\documentclass[runningheads]{llncs}
\usepackage{amsmath}
\usepackage{amssymb}
\usepackage{float}
\usepackage{graphicx}
\usepackage{caption}
\usepackage{subcaption}
\usepackage{array}
\newcolumntype{P}[1]{>{\centering\arraybackslash}p{#1}}
\usepackage[linesnumbered,ruled,vlined]{algorithm2e}

\SetKwInput{KwInput}{Input}                
\SetKwInput{KwOutput}{Output}              
\SetKwInput{KwInitialize}{Initialize}              
\usepackage{algpseudocode} 
\usepackage{xcolor}

\SetCommentSty{mycommfont}
\usepackage{cancel}
\usepackage{bbm}

\begin{document}
\title{RARE: Renewable Energy Aware \\ Resource Management in Datacenters}
%
%
\author{Vanamala Venkataswamy\inst{1} \and
Jake Grigsby\inst{1}
\and \\
Andrew Grimshaw\inst{2}
\and
Yanjun Qi\inst{1}}

\institute{University of Virginia, Charlottesville, VA 22903, USA \and
Lancium Compute, 6006 Thomas Rd Houston, TX 77401, USA
\email{vv3xu@virginia.edu}, 
\email{jcg6dn@virginia.edu},\\
\email{andrew.grimshaw@lancium.com},
\email{yanjun@virginia.edu}}
\maketitle              
\begin{abstract}
The exponential growth in demand for digital services drives massive datacenter energy consumption and negative environmental impacts. Promoting sustainable solutions to pressing energy and digital infrastructure challenges is crucial. Several hyperscale cloud providers have announced plans to power their datacenters using renewable energy. However, integrating renewables to power the datacenters is challenging because the power generation is intermittent, necessitating approaches to tackle power supply variability. Hand engineering domain-specific heuristics-based schedulers to meet specific objective functions in such complex dynamic green datacenter environments is time-consuming, expensive, and requires extensive tuning by domain experts. The green datacenters need smart systems and system software to employ multiple renewable energy sources (wind and solar) by intelligently adapting computing to renewable energy generation. We present RARE (\textbf{R}enewable energy \textbf{A}ware \textbf{Re}source management), a Deep Reinforcement Learning (DRL) job scheduler that automatically learns effective job scheduling policies while continually adapting to datacenters' complex dynamic environment. The resulting DRL scheduler performs better than heuristic scheduling policies with different workloads and adapts to the intermittent power supply from renewables. We demonstrate DRL scheduler system design parameters that, when tuned correctly, produce better performance. Finally, we demonstrate that the DRL scheduler can learn from and improve upon existing heuristic policies using Offline Learning.

\end{abstract}
%
%
%
\section{Introduction}
The sustained demand for digital services has led to record datacenter build-outs and increased energy consumption. Conservative estimates suggest that global datacenter energy consumption between 2010 and 2018 went up by $6\%$, totaling $205$ TWh in 2018. Further research~\cite{Fact_checking} implies that the datacenter energy consumption is an order of magnitude higher than the estimated $6\%$, considering numerous unaccounted small-to-medium scale datacenters and datacenters that cater to new technologies (e.g., blockchain, cryptocurrency mining). Datacenters in the U.S consume $1.8\%$ of the total electricity; electricity predominantly generated using non-renewable sources emitting an estimated $\sim230$ Million Metric tons of greenhouse gases every year.  

Given high carbon emissions and growing societal awareness of climate change, government agencies, non-profits, and the general public demand cleaner (greener) goods and services. Consequently, cloud service providers are investing in green datacenters, i.e., datacenters partially or entirely powered by renewable energy. While some cloud service providers~\cite{DC-google}~\cite{DC-FB} buy carbon offsets, others~\cite{Lancium}~\cite{DC-terascale} are shifting towards datacenters entirely powered by renewables. These datacenters either generate their own renewable energy (self-generation) or draw from an existing carbon-free (e.g., wind, solar) power generation plant (co-location). 

The difficulty with renewables is that power generation is intermittent and subject to frequent fluctuations, making co-location and self-generation interesting from a research perspective. Solar energy generation has a diurnal pattern with maximum energy generation at mid-day, while wind energy generation is higher late in the night (\S \ref{esds}, Fig. \ref{fig:solar_sandia_battery}.b). By combining the solar and wind sources, the energy generation typically complements each other. 

Traditional heuristics-based job schedulers~\cite{Tetrisched}~\cite{Gandiva}~\cite{mesos} use hand-crafted scheduling policies suitable for datacenters with constant power supply. Hand-engineering domain-specific heuristics-based schedulers to meet specific objective functions of highly dynamic green datacenters is time-consuming, error-prone, expensive, and requires domain expertise. A Reinforcement Learning (RL) based job scheduler automatically learns scheduling policies from trial-and-error. The growing body of research~\cite{Google-spotlight}~\cite{DeepEE}~\cite{RLScheduler-SC20} has shown that RL schedulers can learn effective job scheduling policies in traditional datacenter environments with constant power supply. Although the results presented in these works are convincing, they do not examine the complex dynamic green datacenter environments. Furthermore, the existing works treat the RL scheduler as a black box without exploring the design choices (\S \ref{challenges}) that further improve performance. 

Scheduling in a green datacenter encounters additional complexity as the resource pool expands and contracts based on the intermittent and varying power supply. In our previous work~\cite{GreenScheduler-MLCS20}, we demonstrated that the DRL scheduler generates effective scheduling policies, with synthetic power and workload traces, for a cluster of 10 resources. This paper makes the following contributions:
\begin{itemize}
    \item We present a unified green datacenter DRL scheduler, \textbf{RARE}, that allows experimenting with synthetic and real workloads, integrating multiple renewable energy sources and batteries (\S \ref{environment}) to power the datacenter. We show that the DRL scheduler learns effective scheduling policies using synthetic and real HPC workload traces in small and medium scale datacenter environments (\S \ref{syn-hpc-perf}). We demonstrate the DRL scheduler's adaptability to power fluctuations using real power prediction data from renewables (solar and wind) (\S \ref{power-perf}). 
    \item We identified four critical challenges in the existing work (\S \ref{challenges}) and demonstrated performance improvements by appropriately calibrating the DRL scheduler design parameters.
    \item We explore the impact of various DRL scheduler design choices that lead to better performance. Specifically, we demonstrate that the DRL scheduler performs better with a longer planning horizon (in \S \ref{th-perf}) and show the performance implications of choosing the DRL scheduler's neural network configurations (\S \ref{rp-perf}).
    \item We show that the DRL scheduler can learn from and improve upon existing heuristic scheduling policies with Offline Learning techniques (\S \ref{bc-perf}). 
\end{itemize}

\section{Background and Challenges} \label{background}
Job scheduling is deciding when and where to run a set of jobs on a set of resources in order to optimize an objective function. Objective functions define the goal of scheduler optimization. Typical objective functions for schedulers include maximizing revenue for the cloud service provider, maximizing utilization, and minimizing the makespan. The efficient utilization of computing resources leads to millions of dollars in savings for the service providers.

\subsection{Reinforcement Learning (RL) and Job Scheduling} \label{rl-basics}
RL is formalized by a Markov Decision Process (MDP) $\mathcal{M} := (\mathcal{S}, \mathcal{A}, \mathbf{R}, \mathbf{T}, \gamma)$. $\mathcal{S}$ is the set of states, which are representations of information about the environment. $\mathcal{A}$ is the set of available actions that can be taken at each state. $\mathbf{R}$ is the reward function $\mathcal{S} \times \mathcal{A} \times \mathcal{S} \rightarrow \mathbb{R}$. $\mathbf{T}$ is the transition function $\mathcal{S} \times \mathcal{A} \rightarrow \mathcal{S}$ that describes the way actions impact the environment and alter its state. An agent is defined by a stochastic policy $\pi$ that maps states to a distribution over actions in $\mathcal{A}$. A trajectory is a sequence of the states encountered at each timestep, the action taken in those states, and the rewards received $\tau := (s_0, a_0, r_0, \dots, s_T, a_T, r_T)$. The goal of RL is to find a policy that maximizes the discounted sum of rewards over trajectories of experience, denoted $\eta_{\mathcal{M}}(\pi) = \mathop{\mathbb{E}}_{\tau \sim \pi}[\sum_{t=0}^{t=\infty}\gamma^tr_t]$
where $\gamma \in [0, 1)$ is a discount factor that determines the agent's emphasis on long-term rewards instead of short-term outcomes. 

A Deep Neural Network (DNN) is a function approximator that uses layers of nonlinear transformations to learn a mapping between inputs and outputs. The coefficients of a DNN's matrices and vectors are called its weights or parameters. Learning involves finding accurate weights by taking iterative optimization steps along gradient directions that minimize a loss function. Let $\pi_{\theta}$ be a policy parameterized by a neural network with a set of weights $\theta$. $\pi_{\theta}$ takes the current state as input and outputs a distribution over the action space, which can be sampled to make decisions in the environment. 

In this work, we utilize a custom variant of the model-free off-policy actor-critic framework with discrete actions \cite{haarnoja2018soft} \cite{christodoulou2019soft}. The agent interacts with the environment, sampling actions from $\pi_{\theta}$ in state $s$, transitioning to a new state $s'$ and receiving a reward $r$. This experience is saved in a replay buffer $\mathcal{D}$ for later use. In addition to the ``actor network", we initialize a neural network $\phi$ to represent the Q-function, denoted $Q_{\phi}$, which takes state and action vectors as input and outputs an estimate of the expected return when taking action $a$ in state $s$ and following $\pi$ thereafter. We can use our critic network to train the actor network to output higher-value actions. The improved actor is then used to improve the critic network's value estimates, and this process is repeated until performance converges. This technique is ``model-free" because it does not attempt to directly model changes in the environment and ``off-policy" because it recycles data collected from past decisions of the actor network. Further technical details of the implementation are provided in Sec. (\S\ref{drl_agent}).

\subsection{Challenges} \label{challenges}
First, the environment plays a crucial role in RL by providing suitable reinforcement and encouraging the agent to execute the positive actions repetitively. The specially constrained environment rewards or penalizes the agent for correct or incorrect behavior (action). Although existing work~\cite{DeepEE}\cite{RLScheduler-SC20}\cite{Google-spotlight}\cite{RLScheduler-JSSPP2021} has shown RL schedulers learn effective job scheduling policies in datacenter environments (with constant power supply), they do not capture the complex dynamic green datacenters environments where the resource pool expands and contracts (intermittent power from renewables). Additionally, dissimilarities in their environments make it nearly impossible to make a one-to-one comparison among these implementations.

Second, the existing work does not discuss the implications of system design choices, making it difficult to analyze why the RL schedulers perform better than heuristic policies. One such design choice is the size of the planning horizon. The RL scheduler seeks to maximize the future cumulative rewards over some predefined planning horizon. Typically, renewable energy predictions are generated for a $24\mathrm{-}hour$ (day ahead) window. The RL schedulers can make better scheduling decisions with a longer planning horizon, whereas greedy heuristic policies cannot plan for future events. Therefore, studying the DRL scheduler's performance over longer planning horizons is crucial for green datacenters (\S\ref{th-perf}).

Third, the current implementations, discussed in Sec. \S\ref{related-work}, treat the RL schedulers as a black box. These works do not explore RL specific configurations that may significantly contribute to the success of RL schedulers. These configurations may include the neural network size (number of neurons in input, hidden, and output layers) and the state representation (jobs, resources, and power supply). Additionally, following a one-size-fits-all approach while evaluating the RL scheduler with different workloads (with different job properties and size distributions) might diminish the performance. Some of these configuration decisions have performance implications (\S\ref{rp-perf}), while others may influence training time or system memory consumption (not explored in this paper).

Finally, existing RL schedulers overlook the importance of learning and improving upon existing heuristic policies. For instance, designing reward functions that elicit desired behaviors in complex environments is challenging. Instead, the RL schedulers can leverage the behavior of custom heuristic schedulers' designed specifically for unique workloads or environments to learn and improve the overall performance. That is, the heuristic schedulers generate expert demonstrations, and the RL schedulers learn from these demonstrations to improve upon the heuristic policies (\S\ref{bc-perf}). 

\section{Renewable Energy Datacenter Environment} \label{environment}
\begin{figure}[H]
  \centering
  \includegraphics[width=\textwidth]{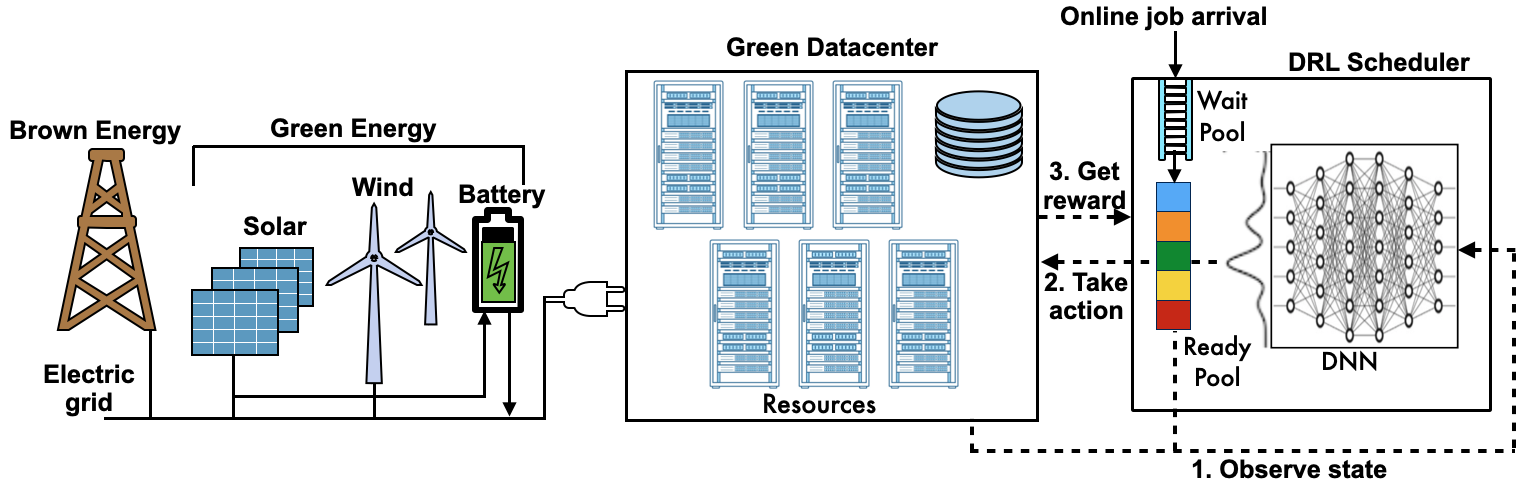}
  \caption{DRL scheduler agent interacting with the green datacenter powered by renewables, battery and electric grid.}
  \label{fig:green-dc-arch}
\end{figure}

The green datacenter is a datacenter co-located at or near renewable energy sources. Various renewable sources can power the datacenter with the provision to store (battery) excess energy from renewables. Additionally, the datacenter is connected to the electric grid to support critical infrastructure when energy from renewables and batteries cannot sustain the load. We aim to design a green datacenter environment that can be controlled by heuristic and DRL-based scheduling policies. In order to train the DRL scheduler agent, we convert the renewable datacenter scheduling problem into an MDP (\S\ref{rl-basics}) with a state space $\mathcal{S}$ describing the current status of the cluster resources, an action space $\mathcal{A}$ of new jobs, and a reward function $\mathbf{R}$ to be optimized. The operation of the datacenter - including receiving new jobs and placing scheduled jobs on available resources - becomes the MDP transition function, $\mathbf{T}$. Fig. \ref{fig:green-dc-arch} provides an overview of a DRL scheduler agent interacting with the green datacenter environment. 

\subsection{State Space, Action Space, and Reward Function} \label{state_space}
The state space, $\mathcal{S}$, includes information about jobs, resources, and resource availability (based on power generation predictions). 

\subsubsection{Resources} 
\begin{figure}[H]
  \centering
  \includegraphics[scale=0.6]{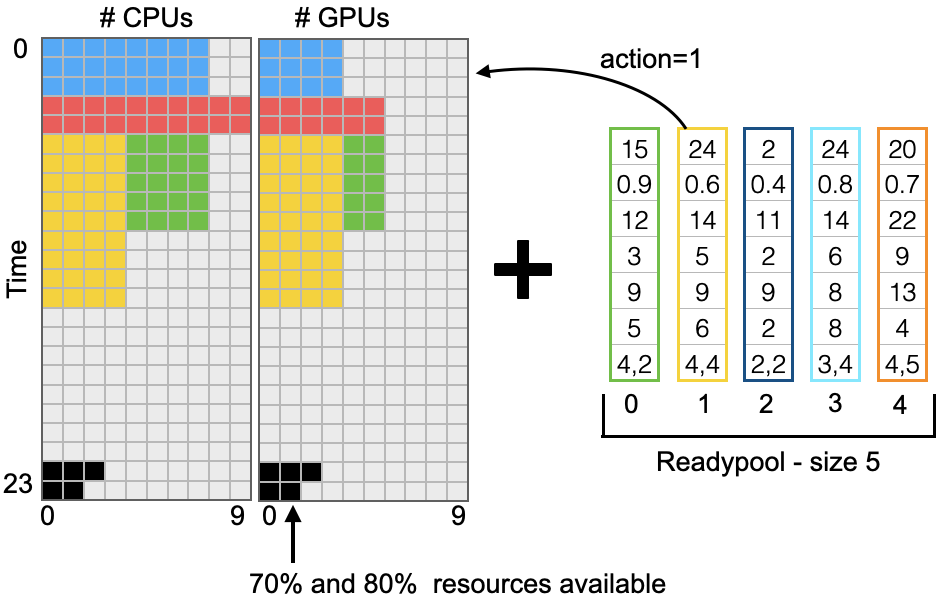}
  \caption{Resource state: A 10 CPUs, 10 GPUs cluster, time-horizon=24, and readypool size=5.}
  \label{fig:resources_jobs}
\end{figure}

Resource availability is represented in an image format of shape $\left(\text{time\_horizon}, \text{resource\_types} \times \text{max\_resources}\right)$, with grey pixels indicating free resources. As jobs are scheduled on the resource pool, segments of the image are occupied by colored rectangles representing jobs' resource requirements and duration. Fig. \ref{fig:resources_jobs} illustrates the cluster image (10 CPUs, 10 GPUs for 24-time units) and the allocation of each resource to jobs scheduled for service, starting from the current timestep and looking ahead 24-time units into the future. Our simulator models a ``pool of resources" (CPUs and GPUs), allowing the scheduler to make granular per resource scheduling decisions. The available resources are allocated contiguously to the jobs (e.g., 8 CPUs and 4 GPUs are allocated to the blue job for three time units).

The power availability feedback is not directly provided to the scheduler agent. Instead, the resource pool expands and contracts based on the power available to the datacenter at any given time. The power availability decides when and how many resources are turned on or off. Therefore, power prediction data is an integral part of the state space, i.e., as power availability changes, the corresponding resource availability is reflected in the state information supplied to the scheduler agent. Resources unavailable due to power constraints are marked black in the resource image. For instance, at timestep 22 (Fig. \ref{fig:resources_jobs}), $70\%$ of power is available, so $70\%$ of the resources are on, and $30\%$ are shut down. Similarly, at timestep 23, $80\%$ of power is available, meaning $80\%$ resources are on and $20\%$ are shut down.

\subsubsection{Jobs} \label {jobs_qos}
In our system, jobs can be in one of three locations: 1) wait\_pool, 2) ready\_pool, or 3) scheduled on the resources. The wait\_pool is where jobs first arrive. The jobs from wait\_pool are moved (in FIFO order) to the ready\_pool where they can then be scheduled on the resources. Jobs have meta-data, including the job's id, value, and resource requirements. The jobs in the ready\_pool are represented as vectors, each job's vector consists of \textit{job\_value, qos, qos\_violation\_time, enter\_time, expected\_finish\_time, duration} and \textit{resource\_requirement}. Additional meta-data for each job is calculated after the job is admitted, e.g., remianing\_runtime (if a job gets suspended) and qos\_violation\_time.

In Fig. \ref{fig:resources_jobs}, the ready\_pool size is 5 (with job indices $0-4$) and has 5 jobs. The yellow job (at ready\_pool[1]) requires 4 CPU and 4 GPU units for the next six timesteps, and the job's value is 24. The jobs are processed over some fixed T timesteps. The time horizon shifts after processing jobs during that timestep, with the job metadata vectors updated and resource image advancing by one row. As the time horizon shifts, the energy available from renewables (and battery) dictates the availability of resources for placing new jobs. The scheduler agent continuously observes the state - consisting of jobs, resources, and resource/power availability - to make scheduling decisions. 

\textit{QoS value}: Users may have different utility functions, i.e., users are willing to pay different amounts for different jobs based on their importance. The user picks the required QoS for that job based on the user's willingness to pay for the job (e.g., spot instances \cite{spot_market}\cite{spot_amazon}\cite{spot_azure}). The QoS value is specified as a percentage of the time the user wants his job to run. The qos\_violation\_time, $\ (expected\_finish\_time \div Qos\ Value$), specifies the upper bound by which the job must finish executing. If a job remains in the system past qos\_violation\_time, it incurs negative rewards every time step after that. The higher the QoS value, the closer the job's completion time to the expected\_finish\_time. If a user wants $0.95 \ (95\%)$ QoS value and specifies expected\_finish\_time $ = 10\ hours$, then the job must be completed within $10.5\ hours$. Expressing QoS value in percentages gives an upper bound of when a user can expect his job to finish. The idea is similar to Least Attained Service (LAS)\cite{gavel} in that if preempted, a job that has received more service is suspended and later restarted.

\subsubsection{Actions} 
The action space for a datacenter with ready\_pool size $n$ is a set of $n+2$ discrete options $\mathcal A = \{j_0, j_1, \ldots ,j_n, suspend, no\_op\}$. The actions $\{a=j_i, \forall i \leq n\}$ schedule the $i$th ready\_pool job $j_i$ on available resources. The job's colored rectangle is added to the first available slot in the resource image with enough free space to schedule it. The action $a=suspend$ is used to suspend an incomplete job and replace it with one of higher value. The suspend action is work preserving, in that a suspended job resumes from the point it was stopped at and not from the beginning. The suspended jobs are re-queued after updating the remaining run time, along with the other ready jobs. Although our scheduler framework supports checkpoint and restart capability~\cite{crac}, the feature was turned off for the experiments discussed in this paper. Finally, the action $a=no\_op$ means that the scheduler agent does not want to schedule (e.g., resources requirements cannot be satisfied) or suspend any jobs in that timestep. In Fig. \ref{fig:resources_jobs}, $action=1$ (at ready\_pool[1]) schedules the yellow job to run on the available resources.

\subsubsection{Rewards} \label{rewards}
The DRL scheduler's objective is realized with rewards that the agent receives. Rewards, which are scalars given by the reward function $\mathbf{R}(s_t, a_t, s_{t+1})$, are a combination of the positive reward or associated cost for the action in a given state. Some actions collect positive rewards, while other actions accrue negative costs. For instance, if a job, $j$, is running on a resource, it collects a positive reward proportional to the job's value. A job's value, $j._{value}$, is calculated based on the type of resources requested, duration, and QoS value. If a job is delayed and QoS violated, it collects a negative reward. Negative reward indirectly encourages fairness, ensuring low QoS value jobs are not delayed or starved. Other costs and rewards can be incorporated into the reward function. Our DRL scheduler's objective is to maximize the total job value from finished jobs, $\vert J_{finished} \vert$ expressed as,
\begin{equation} \label{eq:job_value}
Total\ Job\ Value = \sum_{i=1}^{\vert J_{finished}\vert} j_i.value 
\end{equation}

A direct calculation of value is the price the user is willing to pay to run a job. Total Job Value is both an application-centric and resource-centric metric; the emphasis is on processing as many user jobs as possible, which may increase resource utilization. By processing as many jobs as possible, we essentially maximize the total value we gain from running those jobs. Even a small improvement in total job value can generate millions of dollars in savings for the service providers. Other common objective functions (utilization, makespan, and system throughput) are driven by system-centric parameters that enhance throughput and utilization rather than improving the utility of application processing. These systems treat resources as if they cost the same price and the results of all applications have the same value, even though this may not be the case.

\subsection{Renewable Energy Forecasting} \label{VRE}
Forecasting is crucial for integrating variable renewable energy (VRE) resources such as wind and solar into datacenters. The difference between forecasted output and actual generation is forecast error. Factors that affect forecast performance include forecast time-horizon, local weather conditions, and weather data availability. By integrating VRE forecasts into the scheduling system, datacenter operators can anticipate up- and down-ramps in VRE generation to balance load and generation in intra-day and day-ahead scheduling. 

With shorter timescales, accurate VRE generation forecasting can help reduce the risk of incurring penalties. Over longer timescales, improved VRE generation forecasting based on accurate weather forecasting can help better plan long-running jobs (suspending and resuming the jobs appropriately). The forecasting accuracy decreases with the increase in the forecast time horizon. Thus, selecting a proper time horizon before designing a forecasting model is key to maintaining the accuracy of forecasting at an acceptable level~\cite{forecasting-machine-learning}. 

Additionally, Fig. \ref{fig:green-dc-arch} shows that brown energy can contribute to the electric grid connected to the datacenter. It may not be acceptable to reject or delay some jobs (e.g., jobs with high QoS requirements) if it is possible to execute them with a small additional amount of brown energy. In this case, the datacenter faces a multi-criteria optimization problem comprising the selection of power sources and the scheduling of jobs. Similarly, the use of batteries also results in a multi-criteria optimization problem since the datacenter administrator can decide when to use the additional power from the battery. Multi-criteria optimization is ongoing, and we will cover this topic in our future work.

\subsection{Energy Storage Devices (ESDs)} \label{esds}
ESDs act as a buffer to smooth out intermittent power from renewables, shifting energy from peak generation time-of-day (charging) to low generation periods (discharging). Batteries store the excess energy from wind and solar, increasing the contribution from renewable resources and reducing the electric grid's need. This translates to reduced electricity costs, lower carbon emissions, and highly reliable services. 

\begin{figure}[H]
\centering
\subfloat[$\%$ of time available power exceeds a given value from renewables and battery.]{\includegraphics[height=4.4cm,width=0.49\textwidth]{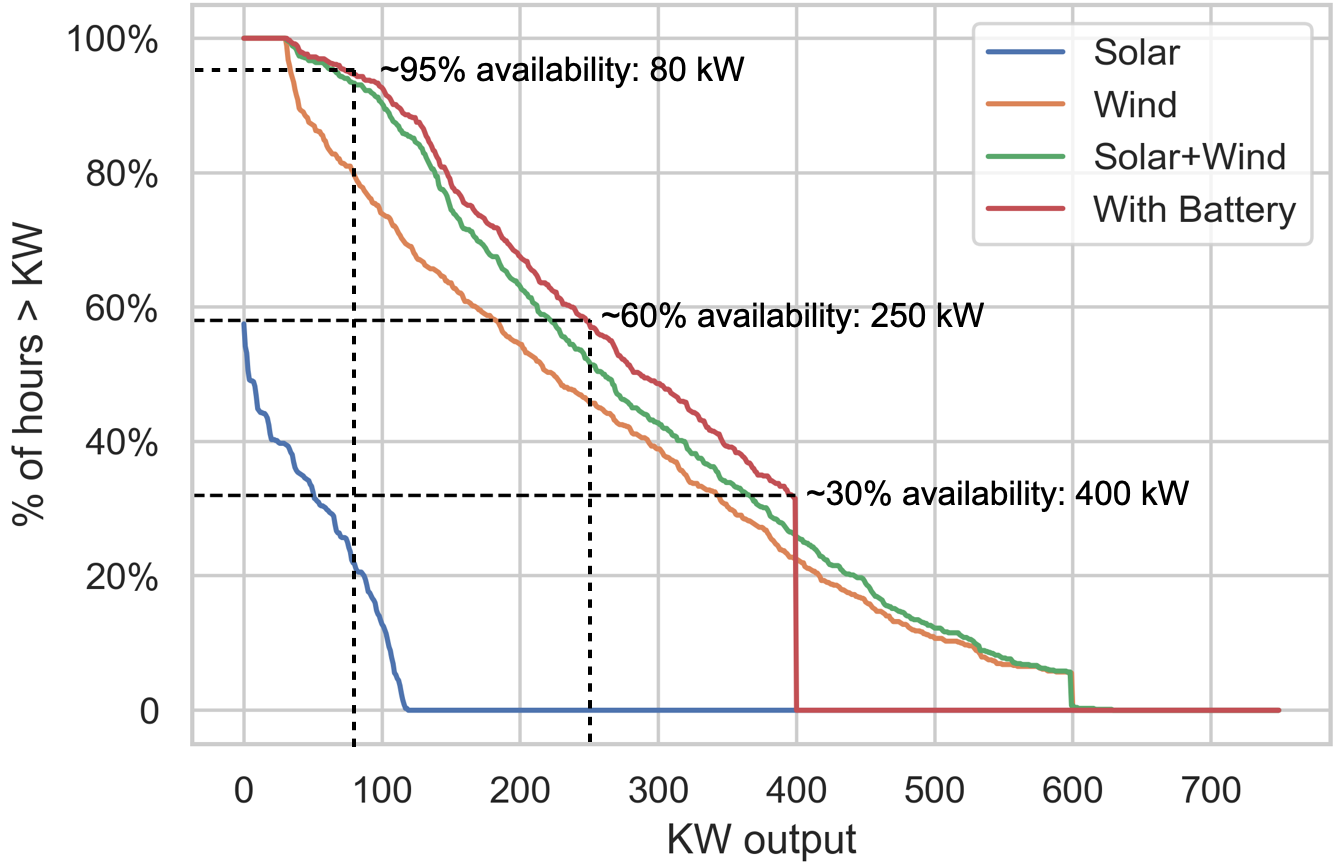}}\hspace{2mm}
\subfloat[Power generation: solar (peak 120 kW), wind (peak 600 kW), June, 2019.]{\includegraphics[width=0.48\textwidth]{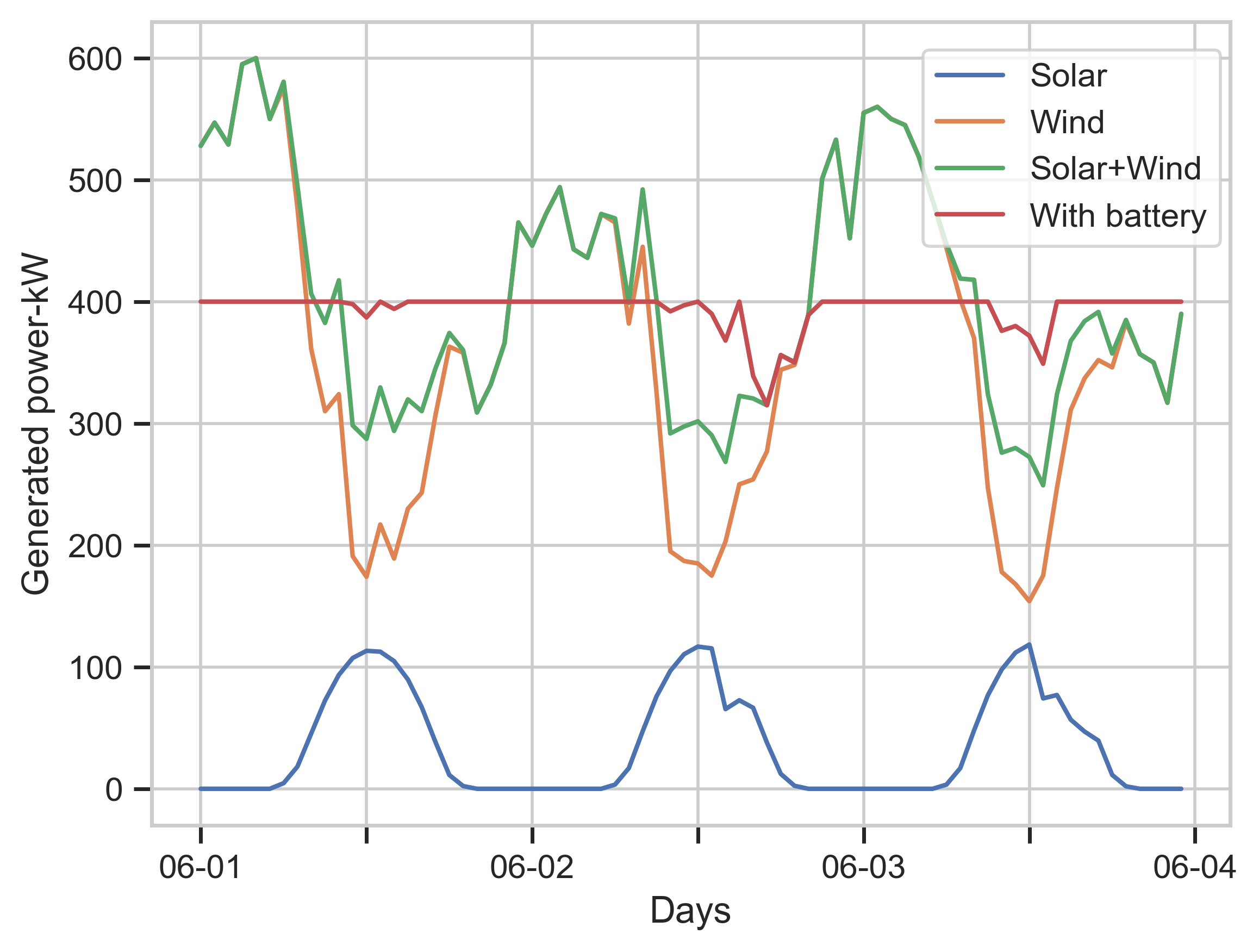}}
\caption{Power generation and power availability from wind and solar}
\label{fig:solar_sandia_battery}
\end{figure}

Fig. \ref{fig:solar_sandia_battery}.a shows the probability that total expected power will exceed a given value for each of renewable sources and combinations from data calculated hourly across the 2019 calendar year at the GLEAMM~\cite{Gleamm} site. For example, a datacenter of 240 kW total electrical draw can expect to have its energy needs met at least $60\%$ of the time entirely by the solar array and wind farm. Approximately 80 kW or more will be available at least $95\%$ of the time for critical infrastructures - such as storage, networking, and control functions of the computational cluster - entirely from the renewables. Under ideal conditions, Fig. \ref{fig:solar_sandia_battery}.b (red line) shows the power (over 3-days in June 2019) from renewables and power available with battery on site. Under normal conditions, the power generation from solar and wind is much more volatile. 

\section{DRL Scheduling Agent} \label{drl_agent}
We train our scheduling agent with a custom variant of the Deep Actor-Critic RL framework (\S\ref{rl-basics}). The datacenter simulator (\S\ref{environment}) with ready\_pool size $n$ is converted to a RL environment that takes an action index $< n + 2$ and returns a new state and reward. States are tuples containing both the resource image and array of job metadata (\S\ref{state_space}), while the reward function can be adjusted to reflect the goals of our scheduling system. This paper focuses on optimizing the total (monetary) value of completed jobs; the reward at timestep $t$ is the total value of all completed jobs at that timestep. Our agent learns to select jobs from the ready\_pool that maximize total job value with the help of three DNNs. 

The \textit{encoder} combines the state information in the resource allocation image and job metadata array and produces a compact vector representation. The resource allocation image is processed by convolutional layers common in computer vision applications, while the job array is passed through standard feed-forward layers.

The two representations are then normalized for stability and concatenated together before a final sequence of layers condenses them to a vector $\Tilde{s} \in \mathbb{R}^{128}$ that summarizes the current state of the scheduling environment. The \textit{actor} network takes $\Tilde{s}$ as input and outputs probabilities for selecting all $n+2$ scheduling actions. The \textit{critic} also takes $\Tilde{s}$ as input and outputs $q \in \mathbb{R}^{n+2}$, where $q[i]$ is an estimate of the total monetary value that we expect to achieve in the future when beginning in the current state and taking the $i$th action.

In an actor-critic method, the actor is trained to assign higher probabilities to actions that the critic determines will lead to higher monetary value. The critic then uses the improved actor to better estimate the expected return of selecting each action. Both networks rely on the state representation $\Tilde{s}$ learned by the encoder network. Separating the encoder in this way lets us share parameters across the actor and critic training processes and reduces overall network size. However, the encoder parameters are updated alongside the critic but not the actor for stability reasons.

Additionally, we are interested in learning to mimic or improve upon the decisions of heuristic schedulers. Learning from fixed datasets of prior experience is the topic of Offline RL \cite{levine2020offline}, and applying standard (``online") actor-critics without the ability to test the policy in the environment and confirm the outcome can lead to value overestimation~\cite{kumar2019stabilizing}. Therefore, we adjust the training process when using offline data so that the actor network learns to mimic the scheduling decisions in the dataset, as long as the critic network suggests those actions are an improvement over what the actor would have done otherwise.

The training process is outlined in Algorithm \ref{algo:training}. Our specific implementation includes several additional details that have been shown to improve stability and performance; the online version of \texttt{RARE} is closest to discrete SAC \cite{haarnoja2018soft} \cite{christodoulou2019soft} while the offline variant is implemented similarly to CRR \cite{wang2020critic}.

\begin{algorithm} [H]
	\SetAlgoLined
	\DontPrintSemicolon
	\KwInput{Batch Size $B$, Learning Rate $\alpha$, \texttt{Online} $True/False$, Discount $\gamma$}
    \KwInitialize{Encoder Net $g_{\psi}$, Actor Net $\pi_{\theta}$, Critic Net $Q_{\phi}$}
    \eIf{\texttt{Online}}{
        \KwInput{Datacenter Simulator Env with Dynamics $\mathbf{T} : \mathcal{S} \times \mathcal{A} \rightarrow \mathcal{S}$ and Reward Function $\mathbf{R} : \mathcal{S} \times \mathcal{A} \times \mathcal{S} \rightarrow \mathbb{R}$}
        \KwInitialize{Replay Buffer $\mathcal{D} \leftarrow \{\}$}
    }{
        \KwInput{Advantage Samples $k$, Replay Buffer with pre-provided transitions $\mathcal{D} \leftarrow \{(s_i, a_i, r_i, s_i'), \dots\}$} 
    }
    \For {training step $t \in \{0,\dots,T\}$}
    {
        \If{Online}{
        \tcp{sample an action from the policy}
            $a_t \sim \pi_{\theta}(g_{\psi}(s_t))$ \\
        \tcp{advance datacenter sim and receive next state and reward}
            $s_t' \leftarrow \mathbf{T}(s_t, a_t),$
            $r_t \leftarrow \mathbf{R}(s_t, a_t, s_t')$ \\
        \tcp{add transition to the replay buffer}
            $\mathcal{D} \leftarrow \mathcal{D} \cup \{(s_t, a_t, r_t, s_t')\}$ 
        }
        Randomly Sample Batch of $B$ transitions $\{(s_i, a_i, r_i, s_i')\}_{i=0}^{i=B} \sim \mathcal{D}$ \\
        \tcp{the encoder embeds the resource image and job metadata into a single array}
        Let $\Tilde{s_j} := g_{\psi}(s_j)$ \\
    \tcp{critic loss (where $\cancel{\nabla}$ cancels gradient contributions)}
    $\mathcal{L}_{critic} \leftarrow \displaystyle \frac{1}{B} \sum_{i=0}^{i=B} \left(\left(Q_{\phi}(\Tilde{s}_i, a_i) - \mathop{\mathbb{E}}_{a' \sim \pi_{\theta}(\Tilde{s}_i')}\left[(r_i + \cancel{\nabla}\gamma(Q_{\phi}(\Tilde{s}_i', a')\right]\right)^2\right)$
    
    \eIf{Online}{
    \tcp{online actor loss (see \cite{haarnoja2018soft}). train the actor to maximize the Q-function.}
    $\mathcal{L}_{actor} \leftarrow \displaystyle \frac{1}{B} \sum_{i=0}^{i=B} \left(\mathop{\mathbb{E}}_{a' \sim \pi_{\theta}(\Tilde{s}_i)}\left[-Q_{\phi}(\Tilde{s}_i, a')\right]\right)$
    }{
    \tcp{estimate the advantage function, $A(s, a)$, by comparing the value of $a$ to the average value of actions sampled from the policy in a given state.}
    Let $\hat{A}(\Tilde{s}_i, a_i) := Q_{\phi}(\Tilde{s}_i, a_i) - \frac{1}{k}\mathop{\Sigma}_{0}^{k}Q_{\phi}(\Tilde{s}_i, a' \sim \pi_{\theta}(\Tilde{s}_i))$ \\
    \tcp{offline actor loss (see \cite{wang2020critic}). supervised regression to copy actions with positive advantage (where $\mathbbm{1}_{\{x\}}$ is $1$ if $x$ is $True$ else $0$)}
    $\mathcal{L}_{actor} \leftarrow \displaystyle \frac{1}{B} \sum_{i=0}^{i=B} \left(-\mathbbm{1}_{\{\hat{A}(\Tilde{s}_i, a_i) > 0\}}\text{log} \pi_{\theta}(a_i | \Tilde{s}_i))\right)$
    }
    \tcp{update neural nets by gradient descent} 
    $\psi \leftarrow \displaystyle \psi - \alpha \nabla_{\psi}\mathcal{L}_{critic}$ , 
    $\phi \leftarrow \displaystyle \phi - \alpha \nabla{\phi}\mathcal{L}_{critic}$ , 
    $\theta \leftarrow \displaystyle \theta - \alpha \nabla_{\theta}\mathcal{L}_{actor}$ , 
    }
    \KwOutput{Trained Scheduling Policy $\pi_{\theta}(g_{\psi}(s))$}
    \caption{\texttt{RARE} Training Process} 
    \label{algo:training}
\end{algorithm}

\section{Evaluation} \label{evaluation}
This section evaluates the DRL scheduler's performance with different workloads, power availability at the green datacenter and explores the effects of design choices on the performance. Before presenting the results, we briefly discuss the workload and experimental setup.

\subsection{Experimentation Conditions}
Our green datacenter simulator compares different resource allocation and scheduling policies using various workloads and power availability settings. Our datacenter simulator (\S\ref{environment}) integrates resources, jobs, power supply from renewables (\S\ref{VRE}) and ESDs (\S\ref{esds}). We have not explicitly modelled networking, storage and I/O overhead. These overheads can be incorporated into the model by adding start/end delay to each job's start and end times. The flexible design of our datacenter simulator allows exploring various design options that can potentially improve the DRL scheduler's performance. For the following experiments, we modeled a small-scale (10 to 50 resources) and a medium-scale datacenter (100 to 300 resources).

\subsection{Evaluation Metrics}
The metric used for evaluating the DRL scheduler's performance is the \textit{Total Job Value} (\S\ref{rewards}) from running the jobs. The \textit{Total Job Value}, accumulated during evaluation, includes a total value for all the jobs that complete on time. The higher the \textit{Total Job Value}, the better. We repeated each experiment 10 times, with new seed, and found the error margin between runs was insignificant.

We also evaluate traditional heuristic scheduling policies, including: Shortest Job First (SJF), Quality of Service (QoS), Highest Value First (HVF), and First Come First Serve (FCFS) for comparison. With the SJF heuristic policy, the job with the shortest runtime is picked first. The job with the highest QoS value (refer \S\ref {jobs_qos}) is scheduled first with the QoS scheduling policy. The highest value job is scheduled first with the HVF policy, and the job with the earliest enter\_time is scheduled using FCFS. Our framework does not support backfilling during scheduling; we will incorporate this feature in our future work.

\subsection{Workload} \label{workload}
The datacenter simulator consists of a cluster with different resource types. Jobs arrive at the cluster in an online manner in discrete timesteps. We assume that the resource demand of each job is known upon arrival; i.e., the resource requirements of each job $j$ is given by the vector $r_j = (r_{j,1}, r_{j,2})$, and $T_j$ is the duration of the job. We assume each job has a fixed allocation (no malleability), such that $r_j$ must be allocated continuously from when the job starts execution until completion. If a job gets suspended, then the job's remaining run time is updated when the job resumes.

\subsubsection{Synthetic Workload} \label{synth-workload}
\hfil\break
We used a synthetic workload where each job consists of meta-data, including job-id, resource requirement (\#cpus, \#gpus), and job duration. Jobs arrive online according to a \textit{Poisson process}. The average job arrival rate, $\lambda$, determines the average load on the cluster. We chose the job duration and resource requests such that $70\%$ of the jobs are short jobs with a duration between $1t$ and $10t$ chosen uniformly. The remaining are long duration jobs chosen uniformly from $10t$ to $30t$ for a time horizon of $48$. Each job can request a maximum of $50\%$ of the total resources, picked randomly. Synthetic workload provides more nuanced control over simulation parameters (e.g., job arrival rate, job distribution) while allowing us to study the scheduler's behavior under a wide range of conditions~\cite{keynote-jsspp}~\cite{synth_workload}.

\subsubsection{HPC Workload} \label{workload}
\hfil\break
We trained and evaluated the RARE scheduler using Argonne National Laboratory (ANL) Intrepid HPC workload~\cite{ANL-intrepid}. The logs contain several months' accounting records ( from 2009) from the Blue Gene/P system called Intrepid. The ANL HPC workload is an old data set, but it has similar characteristics to modern workloads in terms of job arrival rates, resource requirements, and job duration. We made additional changes to the job logs to compensate for missing information. For example, we added GPU requirements to the job requests because Intrepid job logs did not have GPU jobs. Similarly, ANL logs do not have a QoS parameter. We added QoS value (ranging between $0.1$ to $0.9$, refer \S\ref {jobs_qos}) for each job during training and evaluation.

\subsubsection{Power Availability} \label{power-supply}
\hfil\break
We use synthetic power and real power prediction data traces in our experiments. When using synthetic power traces, the power availability level, e.g., $90\%$, means that $90\%$ of the resources are turned on ($10\%$ resources turned off) for that time step (refer Fig. \ref{fig:resources_jobs}). The real power prediction data (solar and wind) is from GLEAMM~\cite{Gleamm} datacenter. The GLEAMM center is a microgrid equipped with 150 kW solar power and three wind turbines connected to the facility, each with 300 kVA of expected power generation. 

\subsection{Results}
First, we evaluate our DRL scheduler with synthetic and HPC workloads. Second, we demonstrate the DRL scheduler's adaptability to the intermittent power supply. Third, we evaluate design choices, namely extended planning horizon and increasing ready\_pool size, that significantly increase the performance of the DRL scheduler compared to heuristics. Finally, we show that the DRL scheduler can learn to imitate the existing heuristic policies and improve performance over those heuristic policies. 

\subsubsection{Performance with Synthetic and HPC Workloads} \label{syn-hpc-perf}
\hfill \break

\begin{figure}[H]
  \centering
  \includegraphics[width=\textwidth]{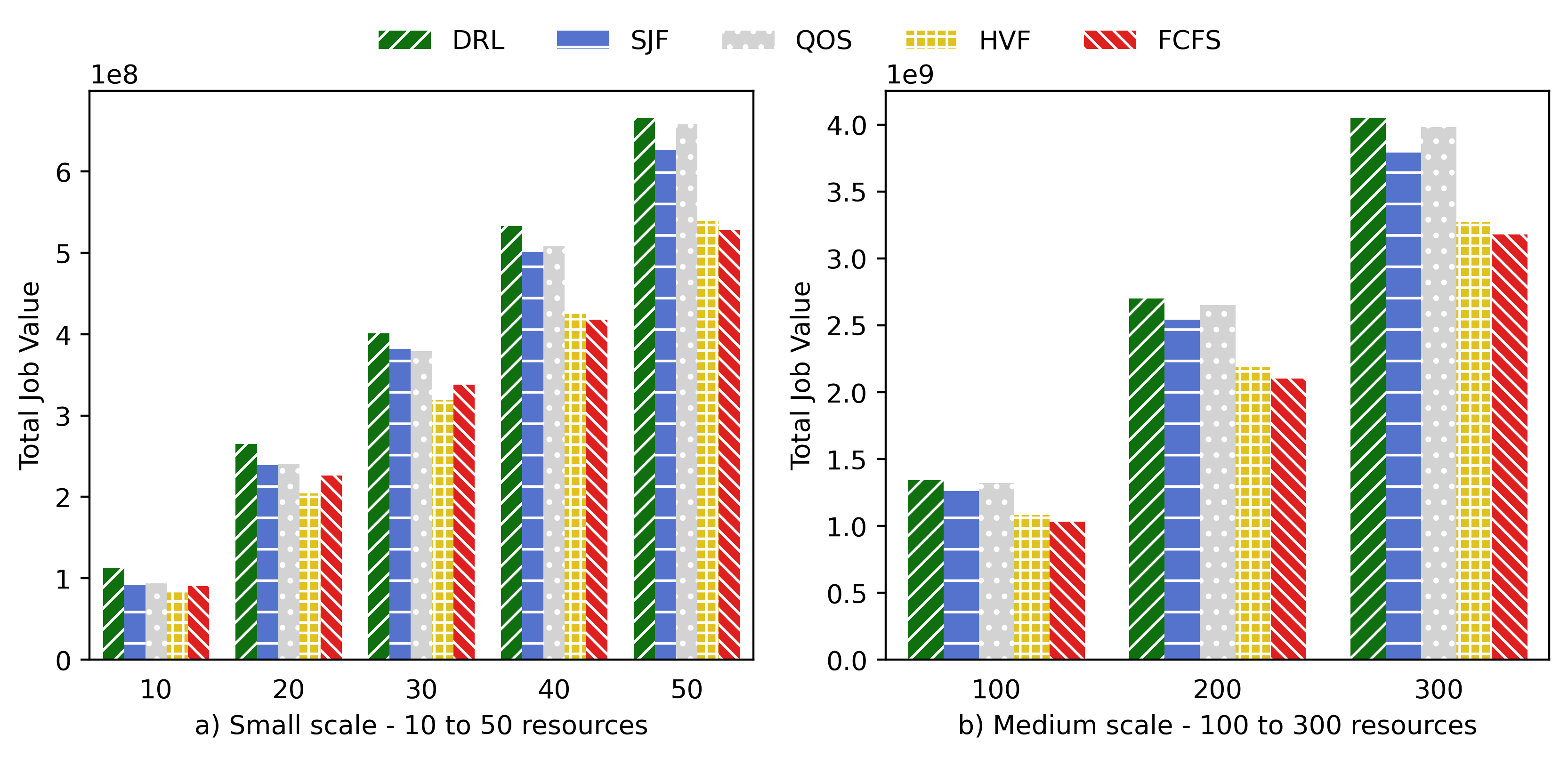}
  \caption{DRL scheduler's performance vs. heuristic scheduling policies with varying resource pool size (small to medium scale datacenter)}
  \label{fig:scaling_perf_comp}
\end{figure}

\textit{Performance with synthetic workload and power data}: This section demonstrates the DRL scheduler's performance with the increasing number of resources. We modeled a small and medium scale datacenter and maintained the workload and power availability at $100\%$. The \textit{Total Job Value} obtained compared to heuristic scheduling policies is plotted in Fig.\ref{fig:scaling_perf_comp}. 

\textit{Analysis}: 
From Fig.\ref{fig:scaling_perf_comp}, the DRL scheduler performs $18\%$ to $25\%$ better for small scale datacenter and $2\%$ to $20\%$ better for medium scale datacenters compared to heuristic policies. As the number of resources increases ($\geq 50$), the DRL scheduler's performance closely matches ($2\%$ to $6\%$ better) the performance of QoS and SJF policies. The DRL scheduler's state space increases as the problem size increases (100 to 300 resources). As the state space increases, the DRL scheduler must explore more states to decide on the best action in any given step. Given the vast state space (for resources $\geq 50$), the agent cannot explore all possible state-action pairs within the fixed episodic limits. Therefore, the performance difference between DRL scheduler and heuristic policies narrows with a larger state space. This huge state-space problem can be alleviated by splitting the state space into smaller sizes. We will investigate this approach in the future.

\textit{Performance with ANL workload and real power prediction data}: We modeled a small scale datacenter (10 to 30 resources) with ANL HPC job workload and maintained the job arrival rate at $100\%$. Additionally, we used the actual power prediction data from the GLEAMM datacenter to simulate a real-world green datacenter powered by renewables and battery (no brown energy).

\begin{figure}[H]
  \centering
  \includegraphics[scale=0.5]{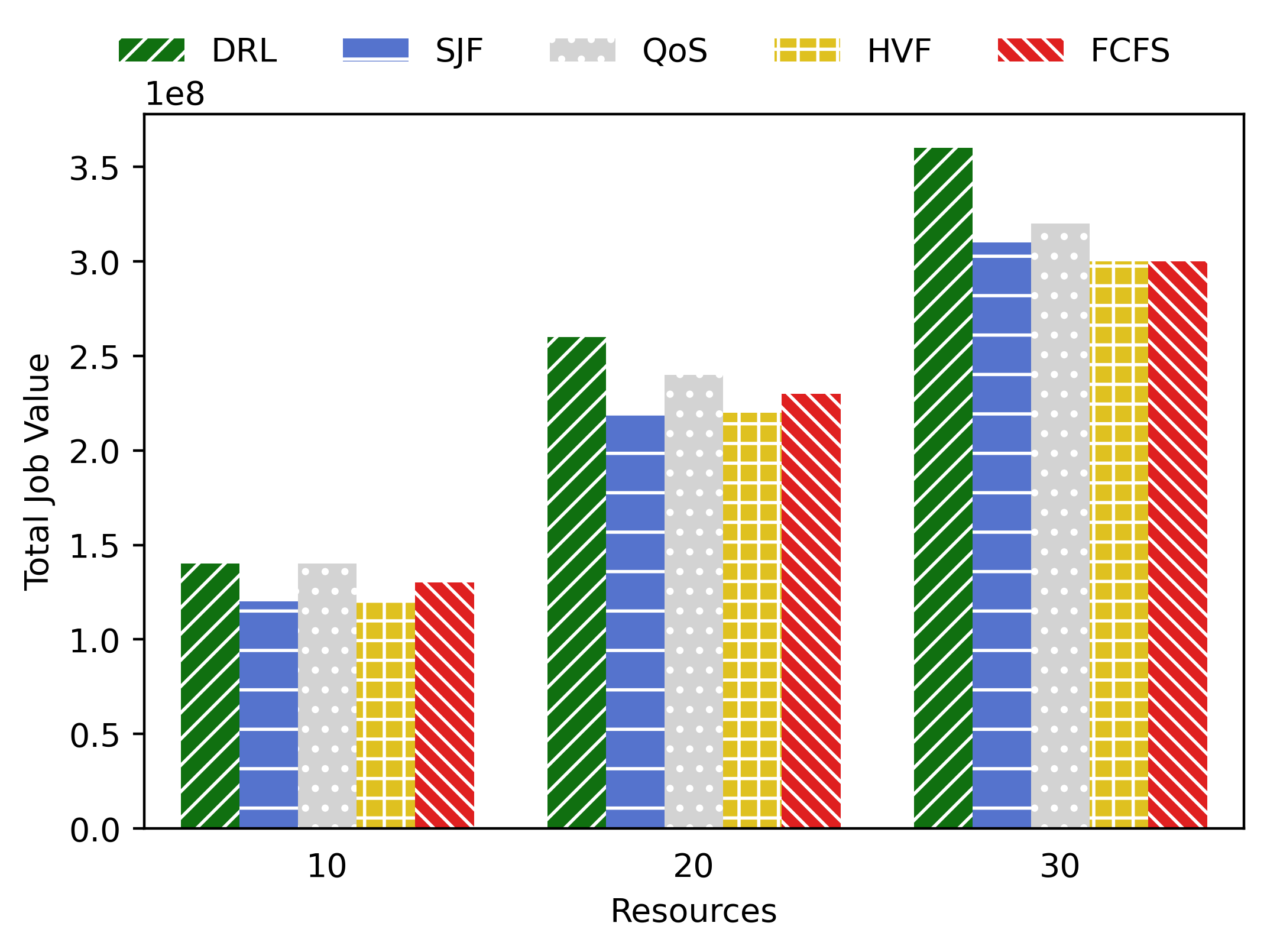}
  \caption{DRL scheduler's performance vs. heuristic scheduling policies with ANL workload and GLEAMM power data}
  \label{fig:hpc_gleamm_workload}
\end{figure}

\textit{Analysis}: Fig.\ref{fig:hpc_gleamm_workload} shows the performance of the DRL scheduler compared to heuristic policies on the ANL workloads and real power prediction data from GLEAMM. For 10 resources, the DRL scheduler's performance matches the QoS and is $5\%$ to $10\%$ better than other scheduling policies. For 20 and 30 resources, the DRL scheduler performs $7\%$ to $14\%$ better than the heuristic policies. 

Different workloads have different job mixes and distributions; therefore, their performance varies~\cite{RLScheduler-SC20}. Although the one-size-fits-all approach works, we plan to investigate further the diverse workload properties to gain deeper insights into designing DRL schedulers (e.g., DNN shape, size, and state representation).

\subsubsection{Scheduler's Adaptability to Intermittent Power Supply} \label{power-perf}
\hfill \break
This section presents the DRL scheduler's adaptability to the varying power supply. The intermittent power generation by renewables necessitates the datacenter resources to switch between power states (off, idle, full throttle). Our experiments simulate intermittent power supply to the datacenter at each time step, not fixed reduced power supply. We modeled small and medium-scale datacenter with different power availability levels and measured the total job value obtained at each level. The resource pool size expands and contracts at every timestep (Fig. \ref{fig:resources_jobs}), based on power availability.The job arrival rate is kept constant at $100\%$ for all the power availability levels. We did not simulate adaptive throttling to dynamically manage the datacenter load since it is out of the scope of this paper.

\begin{figure}[H]
  \centering
  \includegraphics[width=\linewidth]{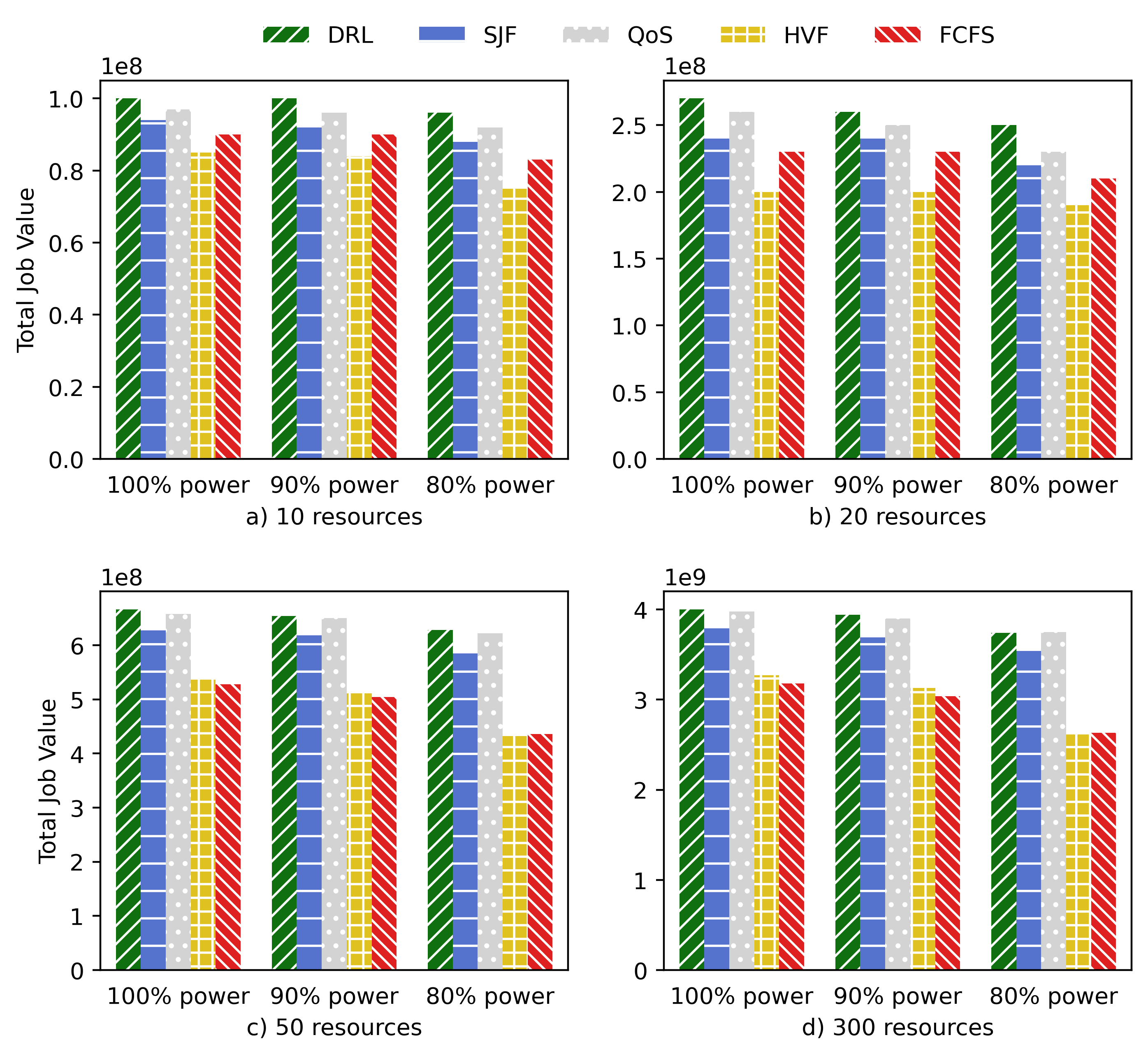}
  \caption{DRL Scheduler's performance with varying power supply - small and medium scale datacenter}
  \label{fig:res_10_20_power}
\end{figure}

\textit{Analysis}: In Fig. \ref{fig:res_10_20_power}, we plotted the total job value with varying power supply ($100\%, 90\%$ and $80\%$) for small and medium size cluster. For small-scale cluster (Fig.\ref{fig:res_10_20_power}.a, b and c), the DRL scheduler performs $9\%$ to $13\%$ better (10 and 20 resources) and $8\%$ to $12\%$ better (50 resources) than heuristic policies. For medium-scale cluster (Fig.\ref{fig:res_10_20_power}.d), the DRL scheduler performs $1\%$ better than QoS policy and $5\%$ to $20\%$ better than other heuristic policies. The greedy heuristic policies, like SJF, do not plan for the future by design. On the contrary, the DRL scheduler observes the resource availability changes in the future and intelligently schedules suitable jobs maximizing total job value. As observed in the previous section, the performance difference between DRL scheduler and heuristic policies narrows (300 resources) with a larger state space.

\subsubsection{Extended Planning Horizon} \label{th-perf}
\hfill \break
Typically, the renewable energy predictions are generated for a $24\mathrm{-}hour$ (day ahead) window. More recently, researchers have developed better prediction models that can predict (with relative accuracy) power generation for extended time windows (2-3 days)~\cite{deepmind-wind}. This subsection investigates the DRL scheduler's performance with various planning horizons, namely 36, 48, 60, and 72 time units. For this experiment, we used synthetic workload and $100\%$ power to isolate the performance implications of the extended planning horizon.

\begin{figure}[H]
  \centering
  \includegraphics[width=\linewidth]{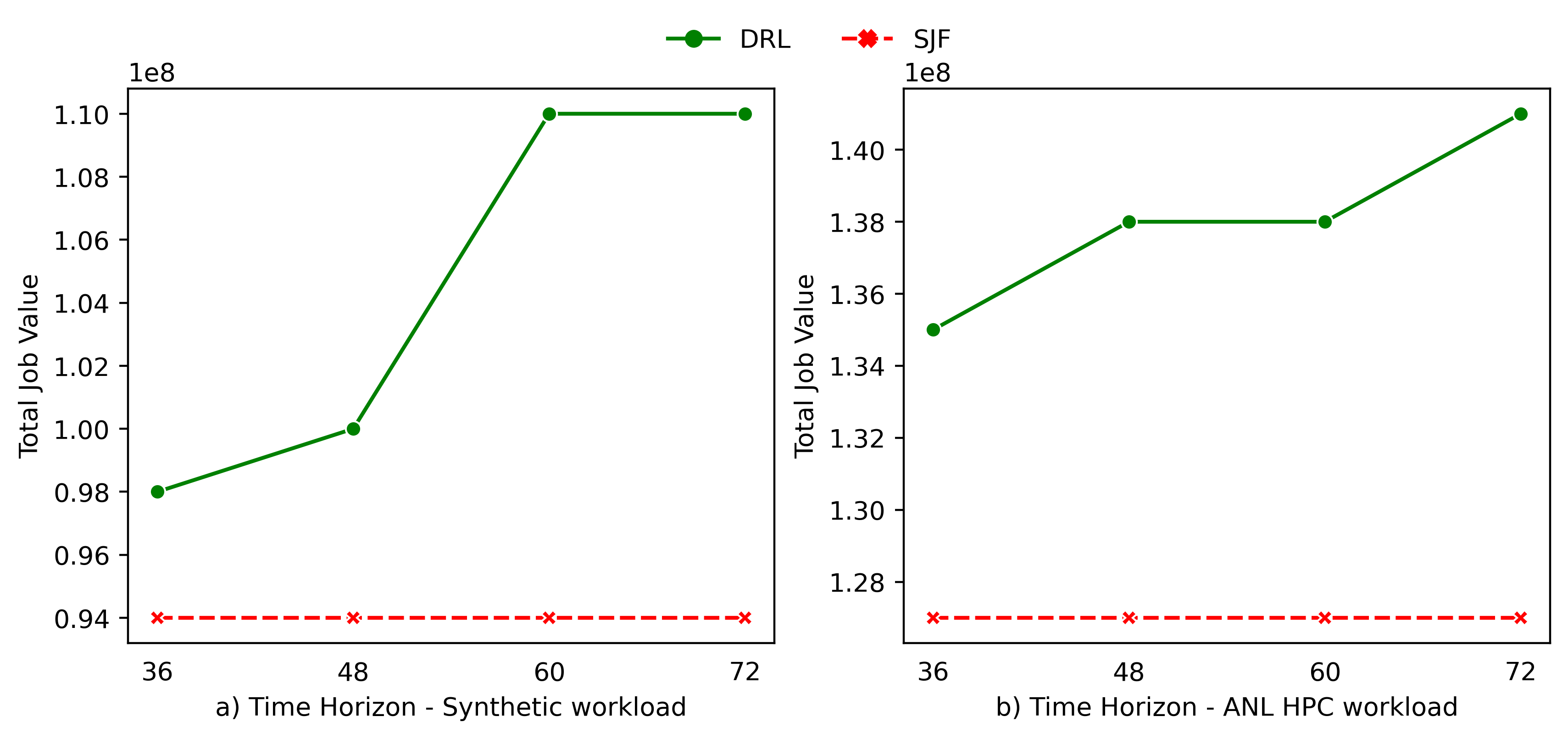}
  \caption{DRL scheduler's performance vs. SJF scheduling policy with increasing time horizon - 10 resources}
  \label{fig:timehorizon_comp}
\end{figure}

\textit{Analysis}: From Fig. \ref{fig:timehorizon_comp}, as the planning horizon increases from 36 to 72, the DRL scheduler performs $4\%$ to $14\%$ and $6\%$ to $10\%$ better than SJF heuristic policy for synthetic and ANL HPC workloads, respectively. The DRL scheduler seeks to maximize the future cumulative rewards over some predefined planning horizon (a.k.a, time horizon). With a shorter planning horizon, TH=36, the DRL scheduler might be limited to \textit{myopic} decisions yielding immediate gains. The greedy heuristic policies lack the ability to plan for future events; specifically, the performance of SJF policy cannot improve as long as the jobs' runtimes are strictly less than the planning horizon. 

Our experiments assume that the quality of predictive information does not decay with an extended time horizon. In reality, as the time horizon increases, uncertainty increases due to weather prediction inaccuracy (described in \S\ref{VRE}). This uncertainty can be captured by changing the $discount\ factor, \gamma$. The discount factor determines how much the DRL agent cares about rewards in the distant future relative to those in the immediate future. If $\gamma=0$, the agent will be completely myopic and only learn about actions that produce an immediate reward. For our experiments above, we set $\gamma = 0.99$. We note that optimization problems become computationally-intensive (due to state-space explosion) with longer time horizons. In the future, we will identify the limits beyond which extending the time horizon will yield ineffective results for the DRL scheduler.

\subsubsection{Varying Readypool Size} \label{rp-perf}
\hfill \break
This section evaluates the performance of different scheduling policies as the ready\_pool size varies. The size of the ready\_pool (described in \S\ref{fig:resources_jobs}) is fixed for any given problem size because the DNN's shape cannot change dynamically during training or evaluation. The DRL scheduler can only select one or more jobs in the ready\_pool at each step. On the other hand, typical heuristic schedulers can select one or more jobs from all of the waiting jobs in the system. This experiment demonstrates that limiting the list of jobs (ready\_pool size) does not affect the performance of the DRL scheduler. For this experiment, we used 10 resources, synthetic and ANL workloads with $100\%$ power supply to study the effect of ready\_pool size on the quality of the results produced by the DRL scheduler.

\begin{figure}[H]
  \centering
  \includegraphics[width=\linewidth]{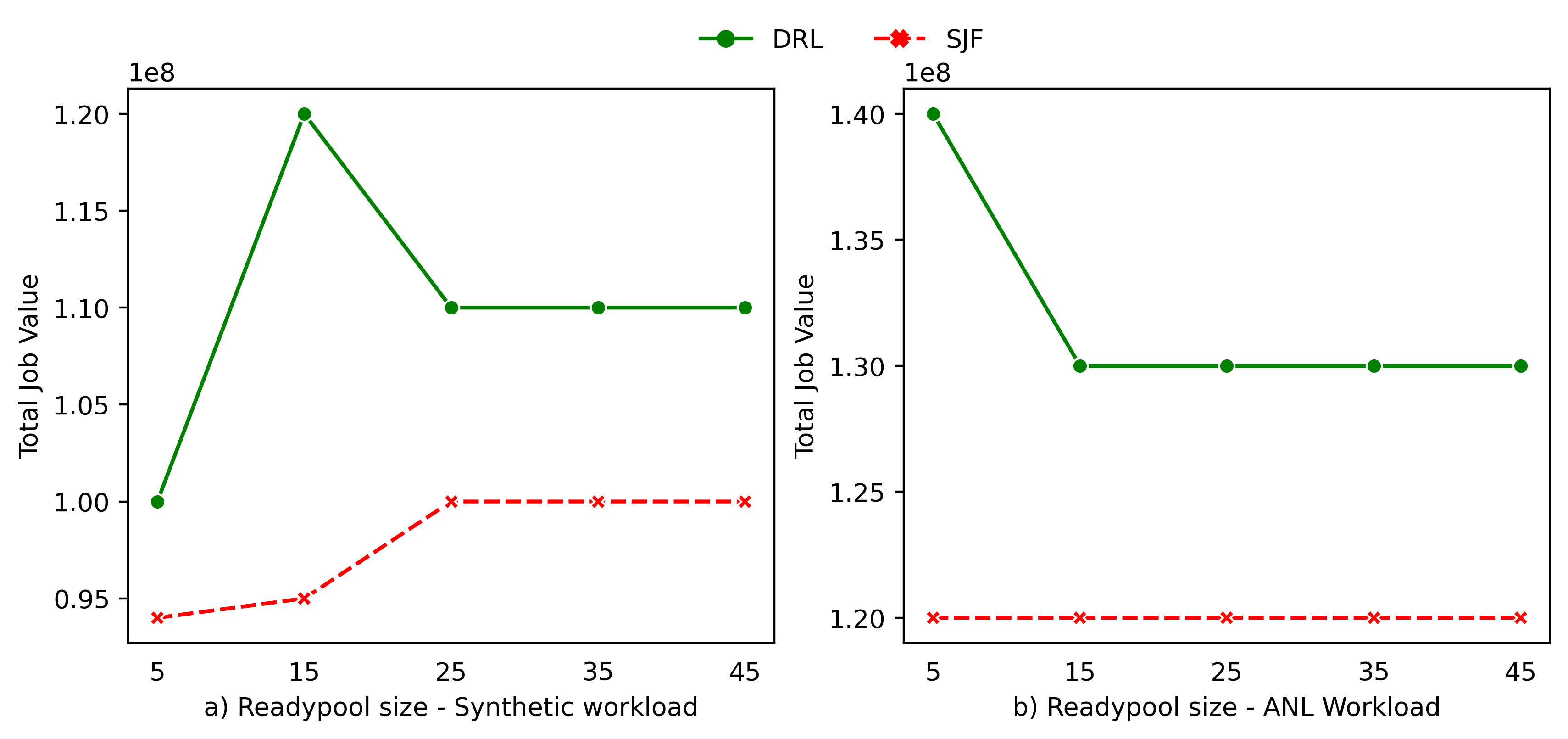}
  \caption{DRL scheduler's performance vs. SJF scheduling policy with varying $ready\_pool$ sizes - 10 resources}
  \label{fig:readypool_sizes}
\end{figure}

\textit{Analysis}: Fig.\ref{fig:readypool_sizes}.a shows the DRL scheduler's performance (synthetic workload) compared to SJF scheduling policy with varying ready\_pool sizes. The DRL scheduler performs best with a ready\_pool size of 15, an $18\%$ improvement over SJF. The DRL scheduler's performance decreases for ready\_pool sizes of 25 and higher but still performs $4\%$ to $7\%$ better than SJF. On the other hand, the SJF policy always picks the smallest job, and the performance stays constant even when we increase the ready\_pool size because the jobs' lengths are within a certain distribution (described in \ref{synth-workload}). Even if more jobs are visible (in the ready\_pool) to the SJF scheduler, the job lengths are likely to be similar. 

Fig.\ref{fig:readypool_sizes}.b shows the DRL scheduler's performance with ANL HPC workload. The DRL scheduler performs $10\%$ better than SJF when the ready\_pool size is 5 and $7\%$ better for ready\_pool size 15 and above. Whereas, with the synthetic workload, the DRL scheduler's performance increases, with ANL HPC workload, the performance decreases as ready\_pool size increases. We showed  that having a smaller set of ready\_pool jobs does not affect the DRL scheduler's overall performance. Further, we believe that the graph trends for the two workloads are different due to the differences in the job distributions. We plan to investigate further with other workloads.

\subsubsection{Behavioural Cloning and Offline RL} \label{bc-perf}
\hfill \break
While DRL can learn efficient scheduling strategies, it often takes hundreds of thousands of timesteps to explore the datacenter environment and adapt from a randomly initialized DNN policy. Following the recent success of Offline RL \cite{levine2020offline}, we can try to learn a scheduling policy from a fixed dataset of prior experience collected from existing heuristic schedulers.

We used a synthetic workload with $100\%$ power supply on a small cluster (20 resources). First, we collected 20 rollouts (sequences of 100-200k samples) of offline experience data where the heuristic scheduling policies select the actions. Second, we load the offline experience into the empty replay buffer during the DRL scheduler's training. The simplest offline learning algorithm is behavioral cloning (BC), where we train the DRL scheduler's actor net to directly mimic the action choices of the heuristic data in its replay buffer. To evaluate the success of the learning process, we simulated new rollouts controlled by the original heuristic and measured the percentage of steps where the BC policy's action is equal to the heuristic's decision in the current state. This \textit{action agreement} metric can be prone to compounding errors because we follow the heuristic even when the DNN would have chosen another action, leading to states that are out of the distribution of the policy $\pi_{\theta}$. However, it provides some insight into our agent's ability to learn heuristic policies. 



\begin{table}[H]
\begin{center}
    \caption{DRL scheduler's action agreement of BC and performance improvement over BC with Offline learning}
    \begin{tabular}{P{0.35\textwidth}|P{0.15\textwidth}|P{0.15\textwidth}|P{0.15\textwidth}|P{0.15\textwidth}}
        \hline
        \textbf{20 Resources} & SJF & QoS & HVF & FCFS \\
        \hline
        Action Agreement (BC) &  $98\%$ & $71\%$ &  $75\%$ & $80\%$ \\
        \hline
        Offline Improvement & $2\%$ & $5\%$ & $19\%$ & $7\%$ \\
        \hline
    \end{tabular}
    \label{table:behavioral_cloning}
\end{center}
\end{table}

\textit{Analysis}: The results for $20$ resource environment are shown in Table \ref{table:behavioral_cloning}. We also experiment with fully offline RL (Algorithm \ref{algo:training} with $\texttt{Online}=\text{False}$), where we use reward information to mimic heuristic actions only when our trained critic network interprets those actions to be an improvement over what the current policy $\pi_{\theta}$ would have done otherwise. This approach improves upon BC in terms of percentage increase in Total Job Value shown in the second row of Table \ref{table:behavioral_cloning}.

Another interesting direction is whether our offline agent can learn from multiple heuristic policies. We collected rollouts from SJF, QoS, HVF, and FCFS on a $50$ resource environment and trained the offline version of \texttt{RARE} on a replay buffer filled evenly with actions from each heuristic. We then measured the action agreement of the DRL policy with each of the original heuristics, and found: SJF ($64$\%), QoS ($7$\%), HVF ($31$\%), and FCFS ($16$\%). This suggests that the offline algorithm learns to favor the SJF heuristic, which performs well in the $50$ resource datacenter (Fig. \ref{fig:res_10_20_power}), and easier to mimic SJF than QoS.


\section{Related Work} \label{related-work}
\subsection{Heuristics Schedulers}
Resource allocation or scheduling has been extensively studied in the literature. Tetrisched~\cite{Tetrisched} is a scheduling system implemented for repetitive analytics jobs in datacenters. Tetrisched plans ahead in time using a Mixed Integer Linear Programming (MILP) constraint solver to optimize job placement. Gandiva~\cite{Gandiva} scheduler, implemented on top of Kubernetes, exploits intra-job predictability (time taken for each mini-batch iteration) to time-slice GPUs efficiently across multiple jobs leading to low job latency. These schedulers do not demonstrate their suitability in green datacenter environments. To address the intermittent power supply from renewables, the existing heuristics schedulers~\cite{batch-edr}~\cite{green-hadoop}~\cite{Krioukov:EECS-2012-13} delay the deferrable jobs until the renewable power is adequate or the electricity price is low before the soft deadline of the jobs expires. Deferring the jobs may lead to poor QoS for the users. Additionally, these implementations use hand-crafted heuristics-based scheduling techniques, and reasoning about these heuristics' interactions is complicated and becomes intractable as the number of variables and heuristics increases.

\subsection{RL Schedulers}
The Spotlight~\cite{Google-spotlight} partitions the agent's neural network training operations onto different devices (CPUs and GPUs) for fast model execution. The RL scheduler in~\cite{Spear-ICDCS19} is designed to minimize the makespan of DAG jobs considering both task dependencies and heterogeneous resource demands. DeepEE~\cite{DeepEE} proposes improving datacenters' energy efficiency by considering the jobs scheduling and cooling systems concurrently. The goal, in~\cite{DeepEE}, is to reduce cooling costs in a datacenter rather than optimize job scheduling. The scheduler in~\cite{RLScheduler-JSSPP2021} implements a co-scheduling algorithm based on an adaptive RL by combining application profiling and cluster monitoring. Smoother~\cite{Smoother-2020} is renewable power-aware middleware. 
This work's primary focus is to provide sustained power to the datacenter with stored energy rather than learning to adapt job scheduling given intermittent power supply.

The RL schedulers discussed above do not have power variability as part of the system's internal state. When we consider power intermittency as part of the system state, it changes the problem setting completely. Additionally, the works discussed above treat the RL schedulers as black boxes without exploring crucial system design parameters that significantly improve overall performance. Further, each of these works is designed for specific environments and workloads and therefore cannot be directly compared with one another or used in other settings. Our implementation is a dynamic system with power variability encoded in the system's internal state. Our DRL scheduler's primary focus is to schedule jobs in green datacenters effectively, not predicting renewable energy production, reducing electricity consumption or carbon emissions. 

\section{Conclusion} \label{conclusion}
Datacenters operate 24X7, guzzling megawatts of electricity, relying heavily on brown energy. Brown energy is expensive and harmful to the environment as brown energy generation releases gigatons of greenhouse gases. Concerns regarding carbon emissions have led organizations to raise the bar by adopting a goal of matching their power consumption with renewable energy.

The difficulty with using renewables to power datacenters is intermittent energy generation, accompanied by inaccuracies in power predictions. The degree of inaccuracy varies from one renewable energy source to another, requiring smart systems and system software to carefully balance and intelligently adapt computing to energy generation. The existing heuristic and RL schedulers are not designed for complex dynamic green datacenters. Additionally, the existing RL schedulers do not explain or explore the system design configurations that lead to better performance with proper tuning.

To address these shortcomings, we propose a unified green datacenter scheduler, RARE, that allows experimenting with synthetic and real workloads and integrates various renewable energy sources along with Energy Storage Devices (batteries). We showed that our DRL scheduler performs better than heuristics-based algorithms in the dynamic green datacenter environment for synthetic and real HPC workloads for a cluster of up to 300 resources. The DRL scheduler adapts exceptionally well to the intermittent power supply (synthetic and actual power prediction data). We demonstrated that accurately tuning the system parameters like planning horizon and ready\_pool size leads to increased performance. Finally, we show that the DRL scheduler can effectively learn from and improve the existing systems using Offline Learning techniques.


\bibliographystyle{splncs04}
\bibliography{paper18_bibliography}

\end{document}